\documentclass[journal]{IEEEtran}

\usepackage{graphicx}
\usepackage[T1]{fontenc}
\usepackage[utf8]{inputenc}
\usepackage{mathptmx}
\usepackage{gensymb}
\usepackage{lineno,hyperref}
\usepackage[dvipsnames]{xcolor}
\usepackage{xcolor}
\usepackage{soul}
\usepackage{bbding}
\usepackage{pifont}
\usepackage{wasysym}
\usepackage{amssymb}

\usepackage{multirow}
\usepackage[noadjust]{cite}

\begin{document}

\title{The Effect of Pitch Distance on the Statistics and Morphology of Through-Silicon Via Extrusion}

\author{Golareh~Jalilvand,~\IEEEmembership{Member,~IEEE,}
        Omar~Ahmed,
        Nicolas~Dube,
        and~Tengfei~Jiang,~\IEEEmembership{member,~IEEE}
\thanks{Authors are with the Department of Materials Science and Engineering, Advanced Materials Processing and Analysis Center at the University of Central Florida, Orlando, FL 32816, USA

Author to whom correspondence should be addressed. E-mail: tengfei.jiang@ucf.edu}}


\maketitle

\begin{abstract}
In this work, we investigated the effect of pitch distance on the statistical variation and morphology of extrusion in Cu TSVs and the underlying mechanisms. Extrusion statistics were obtained from TSV samples with two different pitch distances. A notable increase in the magnitude of extrusion was observed in vias with smaller pitch, yet the extrusion spread was largely unaffected. The morphologies of the extruded vias were characterized and categorized, and finite element analysis was carried out to study the effect of pitch distance on stress and deformation. The results suggested that the overlapping of stress fields from neighboring vias resulted in larger stress in the small-pitch vias, which subsequently led to higher extrusion. The morphologies observed in the extruded vias were related to the operation of different deformation mechanisms under the combined effect of stress and microstructure. The statistical spread of via extrusion, which was similar in both groups of vias, was related to the stochastic nature of the via microstructure. By using a thin cap layer of Ta to suppress the vacancy sources at the via top surface, the adverse effect of the pitch distance was minimized and a pronounced reduction of extrusion was achieved in vias of both pitch distances.
\end{abstract}

\begin{IEEEkeywords}

Through-silicon Via, Extrusion, Pitch Distance, Stress, Statistics, Cap layer.

\end{IEEEkeywords}


\section{Introduction}
\IEEEPARstart{T}{hree}-dimensional integrated circuits (3D ICs) using copper (Cu) filled vertical interconnections, known as through-silicon vias (TSVs), offer promising advantages over the conventional two-dimensional technology in performance, power consumption and device density \cite{Banerjee2001,Knickerbocker2006,Knickerbocker2008}. However, during the via-middle fabrication process as well as device testing and operation, the TSV-based 3D ICs are exposed to multiple thermal cycles that generate considerable thermal stress in the devices due to the differential coefficient of thermal expansion (CTE) between Cu and silicon (Si) \cite{Ryu2011,Rangan2008,slevan2009}. Among the ensuing thermo-mechanical reliability concerns, via extrusion is a particularly important one \cite{jiang2015}. Via extrusion, also known as “Cu protrusion” and “Cu pumping”, manifests as the irreversible upward protrusion of Cu and formation of bumps on the top surface of the vias \cite{JALILVAND2019scripta,okoro2010elimination}. This deformation can be large enough to damage the adjacent structures, in particular the back-end-of-line (BEOL) interconnect layers, and subsequently degrade and even fail the 3D IC device \cite{okoro2010elimination,DeMessemaeker2013,DeMessemaeker2014}. Much work has been done to understand and minimize via extrusion \cite{okoro2010elimination,DeMessemaeker2014,Spinella2016,Smith2015}. Notable attempts to reduce via extrusion include implementing a post-plating annealing and decreasing the via diameter \cite{DeMessemaeker2013,DEWOL2011,Heryanto2012}. Despite of these efforts, a persistent statistical variation in via extrusion has been observed, where the tail of the distribution (e.g. <0.1\% of the population) consists of a small number of vias with very large and detrimental extrusion values \cite{Smith2015,Spinella2016,Jalilvand2017,jalilvand2019ectc,Demas2017}.

The center-to-center spacing of the vias, i.e. the pitch distance (p), is an important design parameter for 3D ICs. While the effect of pitch distance on thermal stress has been studied \cite{PAN2017, an2012pitch}, there is scant literature about the effect of pitch distance on via extrusion and its statistics. Che et al. \cite{che2013pitch} studied the effect of pitch on TSV extrusion with numerical methods only, while a few other studies reported such effect in a very small number of vias \cite{mariappan2012pitch,kee2018epitch}. Understanding the role of pitch distance on the statistical variation in via extrusion is crucial because the overall BEOL reliability of a device containing numerous TSVs will be determined by the weakest link, i.e. the tail of the extrusion distribution. In this work, we statistically investigated the effect of pitch distance on the extrusion behaviors of TSVs. For TSVs with two different pitch distances, the extrusion statistics of 300 vias each were obtained and the extrusion morphologies were analyzed. Scanning electron microscopy (SEM) and focused ion beam (FIB) imaging were implemented to investigate the extrusion morphology and Cu microstructure as well as their correlations. Analysis of the role of pitch on via extrusion was facilitated by thermo-mechanical finite element models. Finally, the effect of a cap layer on reducing the pitch effect and improving the statistics of extrusion was studied and the underlying mechanisms were discussed.

\begin{figure*}[!t]
\centering
\includegraphics[width=0.95\textwidth]{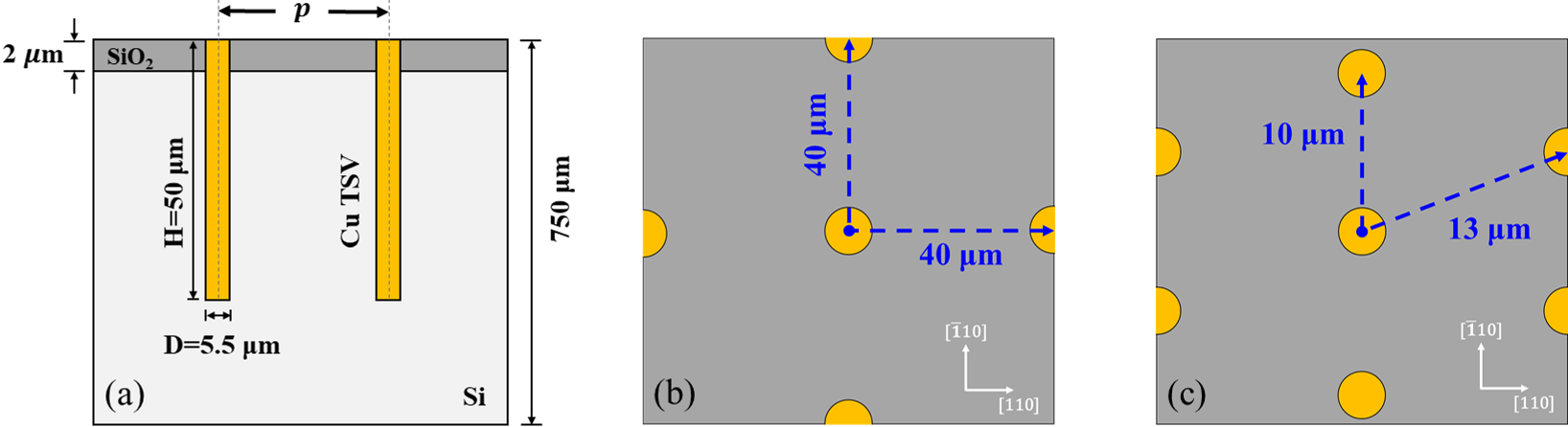}
\caption{(a) Cross-sectional view of the blind via structure. Illustrations of one repeating unit of (b) the large-pitch pattern and (c) the small-pitch pattern.}
\label{fig:1} 
\end{figure*}

\section{Experiments}
\label{experiments}
The TSVs used in this study were blind vias fabricated by a standard via-middle process \cite{GAMBINO2015tsvfab} in a 750$\mu$m thick (001) Si wafer. No post-plating annealing was performed after the overburden was removed by chemical mechanical planarization. The via dimension was 5.5$\mu$m (diameter) $\times$ 50$\mu$m (height) and there was a SiO$_{2}$ layer of 2$\mu$m on the wafer surface surrounding the Cu vias, as illustrated in Fig. \ref{fig:1}a. There was also a 0.4$\mu$m oxide liner and a 0.1$\mu$m Ta barrier at the via side wall. Two via patterns with different via pitch distances, p, were fabricated in the same wafer. The large-pitch pattern consisted of a regular square arrangement of vias with p= 40$\mu$m along the $[\overline{1}10]$ and $[110]$ directions (Fig. \ref{fig:1}b). The small-pitch pattern consisted of a distorted hexagonal arrangements of vias with the smallest pitch of p=10$\mu$m along the $[\overline{1}10]$ direction (Fig. \ref{fig:1}c). A sample containing both patterns was cut from the wafer and dipped into acetic acid to remove any native oxide that might have formed on the surface of the Cu, followed by ultrasonically cleaning with acetone, isopropyl alcohol and deionized water. To induce extrusion, annealing was carried out at 400\degree C for one hour in a forming gas (Ar-4\%H$_2$) atmosphere. The heating rate was 6\degree /min and the sample was furnace cooled to room temperature.

After annealing, the sample was cut into two halves that contained both patterns. On one half, 150nm thick Aluminum film was deposited to enhance the surface reflectivity and the surface morphology of 300 vias for each pattern was measured by white light interferometry (WLI). All the vias measured and used in the statistical analysis were located away from the free edge of the sample. The extrusion morphology was analyzed using the Gwyddion modular program \cite{NecasGwyddion} and the average extrusion and maximum extrusion of each via were obtained using a method described elsewhere \cite{JALILVAND2019scripta}. Since maximum extrusion is the most pertinent to device reliability, analysis and discussion in the rest of the paper would be focused on the maximum extrusion. On the other half of the sample, for selected vias, after imaging by scanning electron microscopy (SEM), the extruded Cu was removed by focused ion beam (FIB) with the ion beam aligned parallel to the via top surface. Low current ion beam milling was used to ensure the removal of less than 100nm of Cu below the original via surface prior to annealing. After milling, low current FIB secondary electron images were obtained for the vias, where grains with different orientations showed different contrasts due to the ion channeling effect \cite{kempshall2001ionchanneling}.

\section{Results}
\label{Results}
\subsection{Extrusion statistics}
\label{Results:statistics}

\begin{figure*}
\centering
\includegraphics[width=0.9\textwidth]{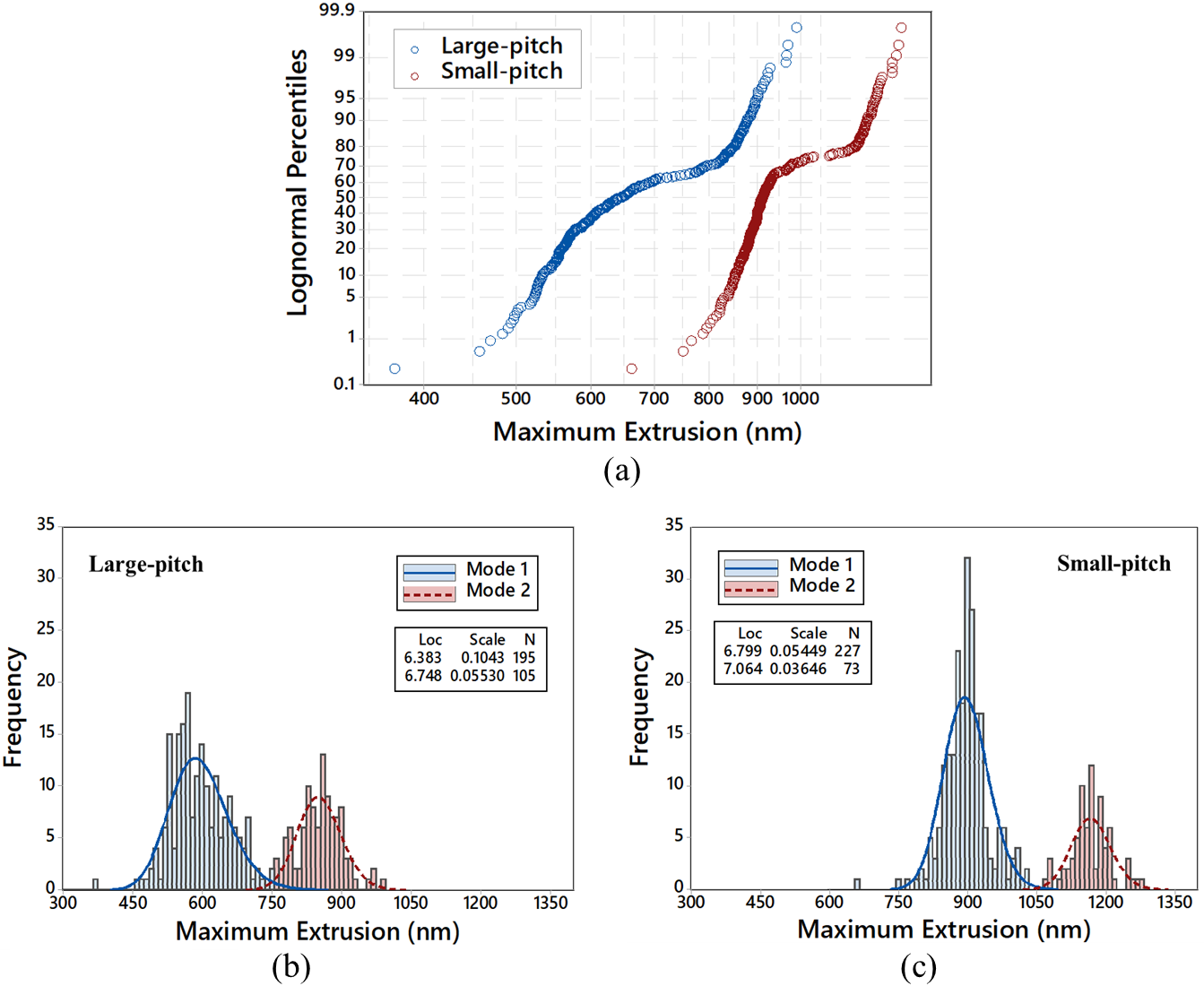}
\caption{(a) Lognormal percentile of the maximum extrusion for 300 large-pitch vias and 300 small-pitch vias. Histograms with fitted lognormal probability density functions for (b) large-pitch and (c) small-pitch vias.}
\label{fig:2} 
\end{figure*}

The lognormal percentile plots of the maximum extrusion for both pitches were presented in Fig. \ref{fig:2}a. A wide spread of extrusion values was observed regardless of the pitch distances. Both curves exhibited a bimodal behavior, where a kink at around $70^{th}$ percentile separated each curve to two segments of similar slopes, and there was an average shift of the small-pitch curve to higher values. At 618nm vs. 613nm, the extrusion range was slightly larger in the large-pitch sample than in the small-pitch sample, although the difference was minimal. The bimodal distributions were further shown in Fig. \ref{fig:2}b and \ref{fig:2}c, where histograms with fitted lognormal probability density functions were plotted for each group of vias. The lognormal distribution location parameter (Loc), scale parameter, as well as the sample size (N), were shown for each mode in the inserts of Figs. \ref{fig:2}b and \ref{fig:2}c. The effect of smaller pitch distance in increasing extrusion was quantitatively seen by comparing the location parameters, which showed that mode 1 and mode 2 of the small-pitch vias were shifted by 305.2nm and 316.8nm from their large-pitch counterparts. Comparing the scale parameters suggested that as pitch decreased, the two extrusion modes were slightly more peaked and less spread out in the small-pitch sample. In both samples, mode 1 contained a larger number of vias, especially in the small-pitch sample. 

\begin{figure}[!b]
\centering
\includegraphics[width=0.45\textwidth]{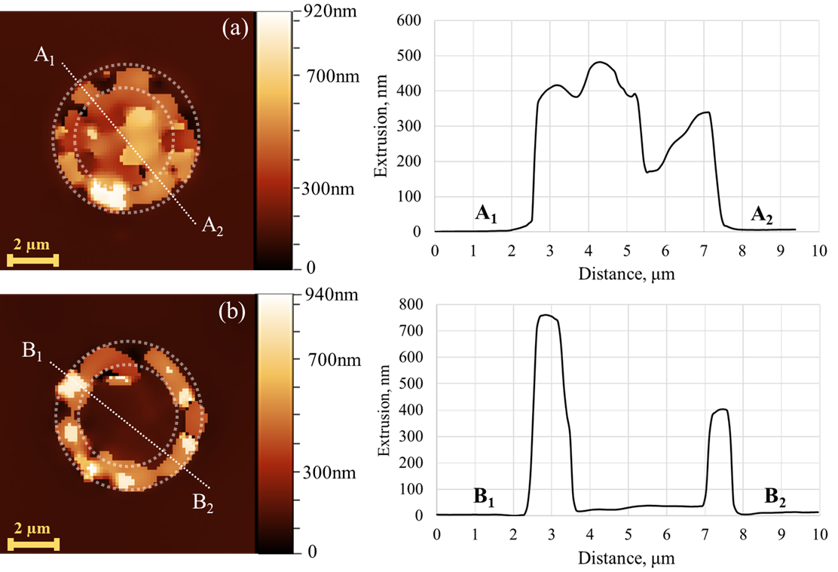}
\caption{(a) WLI surface profile and the height profile along $A_1-A_2$ in a “granular” via. (b) WLI surface profile and the height profile along $B_1-B_2$ in an “annular” via.}
\label{fig:3} 
\end{figure}

\begin{figure*}
\centering
\includegraphics[width=1\textwidth]{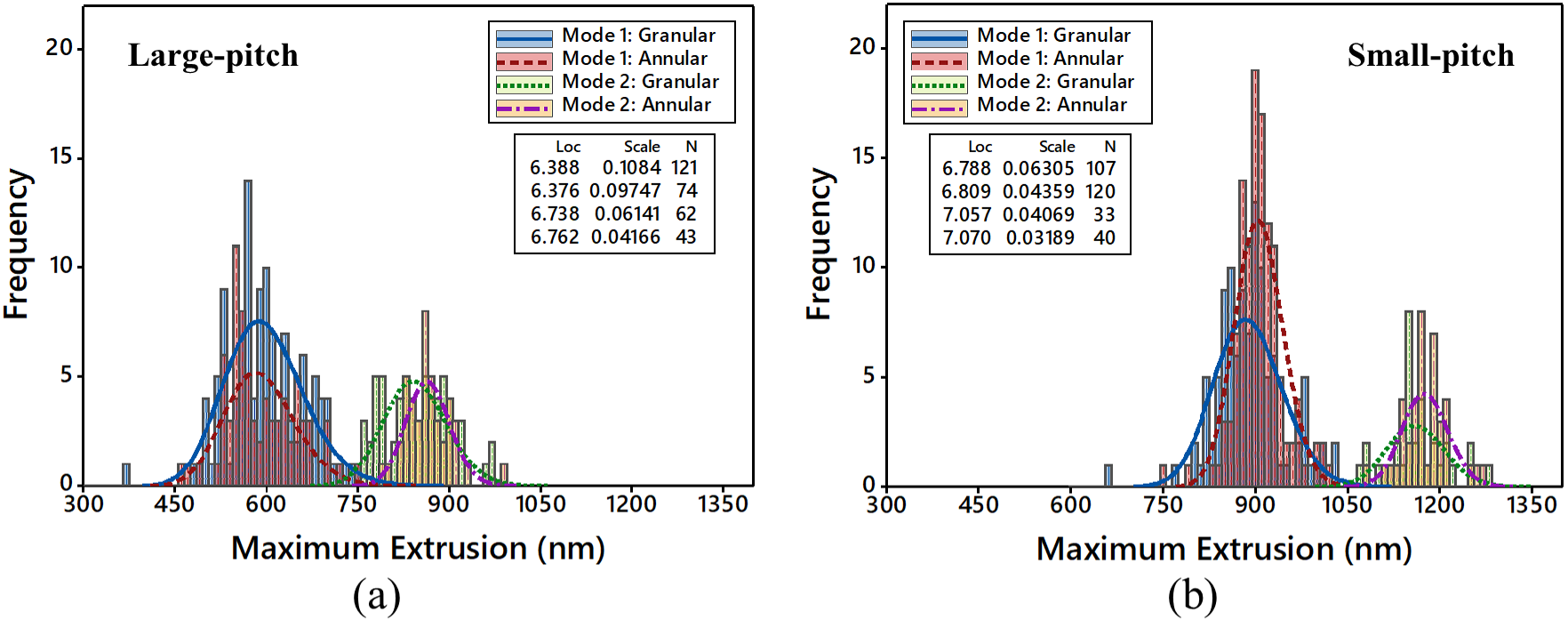}
\caption{Histograms of the granular and annular morphologies in each extrusion mode for (a) large-pitch and (b) small-pitch vias.}
\label{fig:4} 
\end{figure*}

\subsection{Extrusion morphology}
\label{Resulst:morphology}
Examining the extrusion morphologies of the 600 vias measured by WLI revealed a variety of surface features. In some vias, the extruded Cu bumps were almost exclusively at the peripheral of the vias, and in other vias, extrusion height varied across the entire via. To classify the extrusion morphologies, the via surface was divided into two regions with equal area (50\% each), containing a concentric inner region and the annulus. Based on how much extrusion fell in each region, the extrusion morphologies were classified into two categories. In the first category, referred to as “granular” morphology, the inner and outer 50\% of the vias exhibited comparable average extrusions regardless of where the maximum extrusion occurred. In the second category, referred to as “annular” morphology, the maximum extrusion occurred in the outer 50\% of the via surface and the average extrusion of the outer region was more than 5 times that of the inner region. A representative WLI image for each morphology was shown in Fig. \ref{fig:3}a and \ref{fig:3}b, respectively, where the dashed lines marked the circumferences of the vias and the boundaries between the inner and outer regions.

Varying the pitch distance changed the population and distribution of each extrusion morphology. To visualize this effect, distribution histograms of the granular and annular morphologies in the large-pitch and small-pitch samples were shown in Fig. \ref{fig:4}, where the fitted lognormal distribution curves were superimposed. The location parameter (Loc), scale parameter, and the sample size (N) for each morphology in each mode of the two samples were shown in the inserts. Comparing the location parameters indicated quantitatively that the extrusion of the two morphologies in each mode were shifted to larger values when the pitch distance was reduced. Regardless of the morphology and mode, the scale parameter was smaller in the small-pitch vias, indicating tighter extrusion distributions, although the difference was small. Furthermore, the scale parameters in both small- and large-pitch samples showed that the spread of extrusion for the granular morphology was slightly larger in both modes. In the large-pitch sample, there were 183 vias with the granular morphology, making it the dominant morphology. Granular morphology was also the dominant morphology in each individual mode in the large-pitch sample. When the pitch distance decreased, there was a notable increase of vias with annular morphology in the small-pitch sample and  a corresponding decrease of the granular shaped vias (140 granular vs. 160 annular). Annular morphology was also the dominant morphology in each individual mode of the small-pitch sample.

\section{Discussion}
\label{Discussion}
\subsection{The effect of pitch distance on stress}
\label{Discussion:stress}
The expansion of Cu at high temperatures, when constrained by the surrounding Si, subjected the Cu via to large compressive stress, which drove Cu extrusion. When the pitch distance decreased, the stress was enhanced due to the superposition of stress fields from neighboring vias. Using finite element analysis (FEA), the stress and deformation in the Cu vias was examined for the two pitch distances. 3D FEA models were constructed for the small-pitch and large-pitch patterns using the linear 3-D solid elements (C3D8R) in a commercial package, $ABAQUS^{\circledR}$. Symmetric boundary conditions were applied to simulate the periodicity of the patterns, while the upper surface was traction free and the bottom was fixed. The thin barrier and liner layers at the via side wall were ignored in the model. All materials were assumed isotropic with $E_{Cu}=110 GPa$, $\nu_{Cu}=0.35$, $\alpha_{Cu}=17 ppm/K$, $E_{Si}=130 GPa$, $\nu_{Si}=0.28$, $\alpha_{Si}=2.5 ppm/K$, $E_{SiO_2}=72 GPa$, $\nu_{SiO_2}=0.16$, and $\alpha_{SiO_2}=0.55ppm/K$. 

\begin{figure*}
\centering
\includegraphics[width=0.8\textwidth]{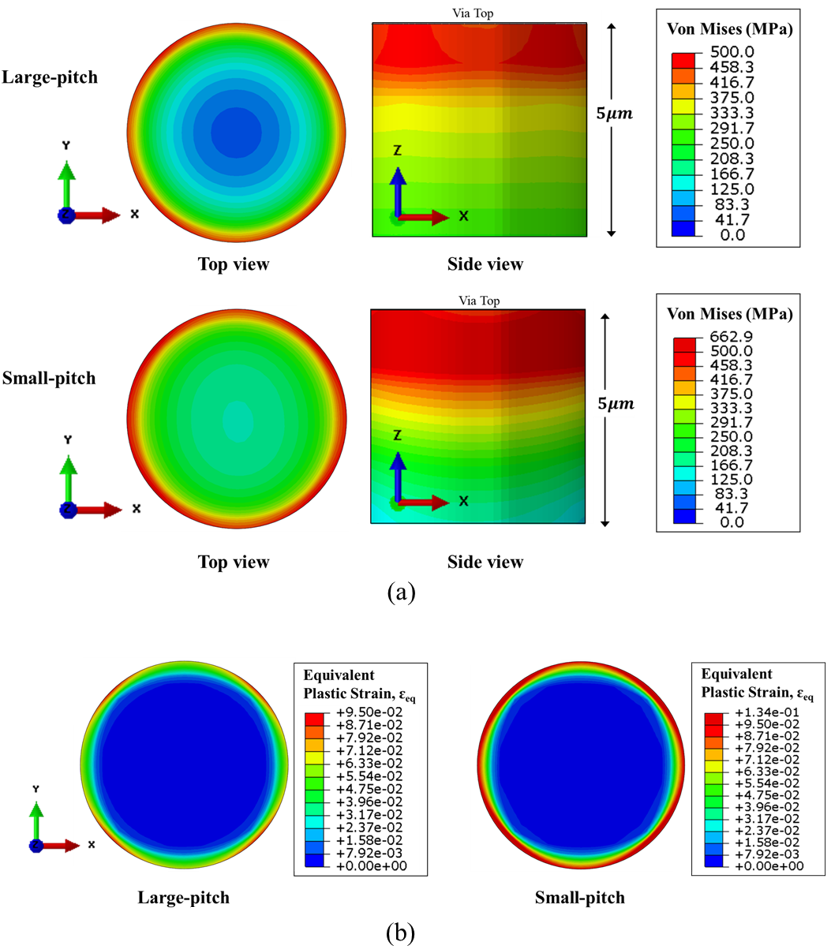}
\caption{(a) Top view and side view von-Mises stress contours for a large-pitch via and a small-pitch via; (b) Equivalent plastic strain for a large-pitch via and a small-pitch via.}
\label{fig:5} 
\end{figure*}

A thermal load of $\Delta T =375$\degree C was considered, which corresponded to heating from room temperature to 400\degree C. First, an elastic model was used to examine the distribution of the von-Mises stress, $\sigma_{v}$, which was the effective shear stress driving plastic deformation \cite{JIANG2013}. In Fig. \ref{fig:5}a, for a via in each pattern, the calculated von-Mises stress contours were shown for the via top surface (top view) and the top 5$\mu$m of the via (side view). In both cases, larger von-Mises stress was clearly seen at the via peripherals and near the top surface due to the triaxial stress state in those locations. The von-Mises stress level was elevated in the small-pitch via due to the overlapping of stress fields from neighboring vias. Using a plastic model with the hardening behavior of Cu described in Table \ref{tab:1} \cite{liu2009failure}, the equivalent plastic strain, $\varepsilon_{eq}$, was calculated for a via in the two via patterns. Shown in Fig. \ref{fig:5}b, larger $\varepsilon_{eq}$ was observed in the top of the small-pitch via, especially along the peripheral of the via. These results showed that reducing the pitch distance led to higher von-Mises stress to drive more plastic deformation in the small-pitch via and the effect was more pronounced along the via peripheral. In an actual via, both microstructure and time dependent deformation processes played important roles in plastic deformation. Although these factors were not considered in the FEA models, the results were still useful in providing a general depiction of the role of reduced pitch distance in increasing stress, particularly near the via peripheral, which in turn promoted plastic deformation to enhance Cu extrusion. 

\begin{table}
\centering
\caption{Plastic properties of Cu used in FEA \cite{liu2009failure}.}
\label{tab:1}       
\begin{tabular}{lll}
\hline\noalign{\smallskip}
Stress (MPa) & Strain   \\
\noalign{\smallskip}\hline\noalign{\smallskip}
120 & 0  \\
121 & 0.001  \\
186 & 0.004  \\
217 & 0.01  \\
234 & 0.02  \\
248 & 0.04  \\
\noalign{\smallskip}\hline
\end{tabular}
\end{table}

\subsection{The effect of pitch distance on the extrusion behaviors}
\label{Discussion:statistics}
The stress demonstrated in FEA drove plastic deformation near the top of the via to manifest as via extrusion. Under the test conditions, several deformation mechanisms were known to be operative in the TSVs, including dislocation glide and climb, creep and grain boundary sliding accommodated by grain boundary diffusion, and interfacial sliding \cite{DeMessemaeker2014,JALILVAND2019scripta,Demas2017,Dutta2019mechanism}. Depending on the microstructure of the via and the stress it was subjected to, one or more mechanisms could dominate to result in the observed morphology and height variations. 

\begin{figure*}[!t]
\centering
\includegraphics[width=0.8\textwidth]{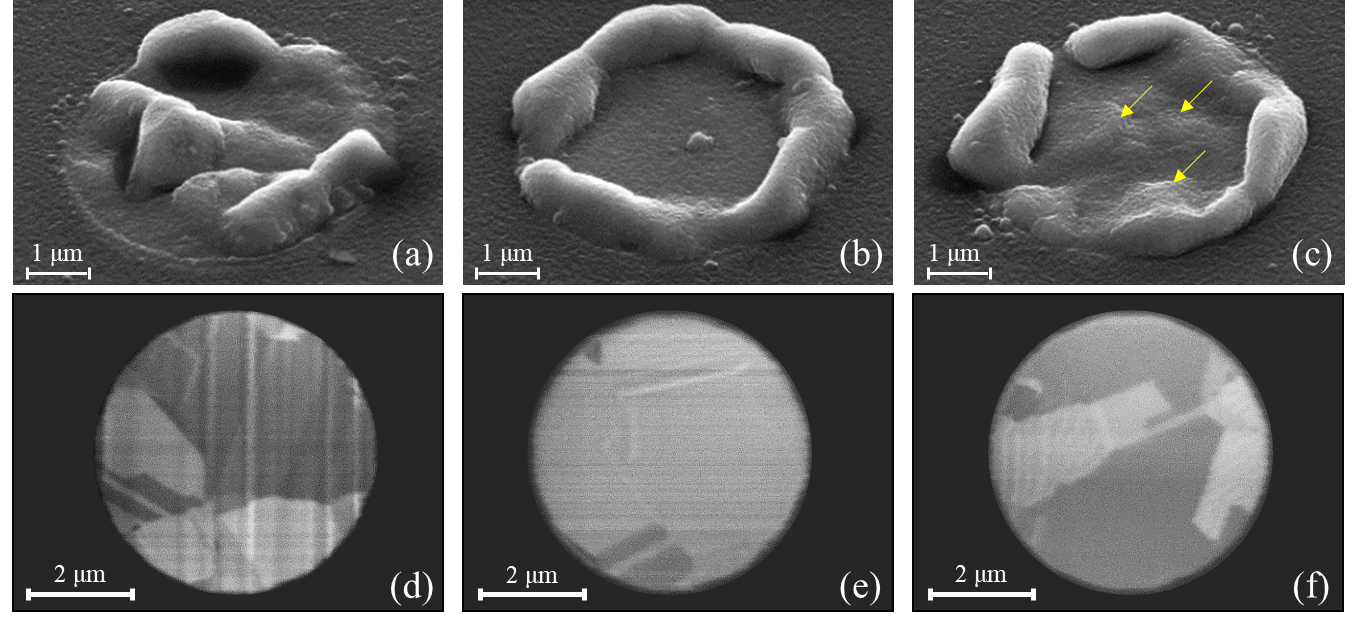}
\caption{(a-c) SEM images at 45\degree tilt of a via with mode 1 granular morphology at 863nm extrusion, a via with mode 1 annular morphology at 809nm extrusion, and a via with mode 2 annular morphology at 1260nm extrusion. (d-f) Corresponding FIB images of the vias in (a-c). Note that the vertical and horizontal lines in (d) and (e) were artifacts from FIB milling and imaging.}
\label{fig:6} 
\end{figure*}

For two small-pitch vias with similar maximum extrusion values of 863nm and 809nm that belonged to mode 1 of the extrusion statistics, SEM revealed that one via had granular morphology with distinct local bumps (Fig. \ref{fig:6}a) and the other via had annular morphology, where the protrusion was relatively uniform at the via peripheral, connecting to a ring (Fig. \ref{fig:6}b). The corresponding FIB grain contrast images in Figs. \ref{fig:6}d and \ref{fig:6}e qualitatively revealed very different grain structures at the top surface of these two vias. In the via in Fig. \ref{fig:6}a, several grains and high density of grain boundaries could be seen in both the inner and outer regions (Fig. \ref{fig:6}d). The grain contrasts suggested the possible existence of grain boundaries with large misorientation angles. It was known that the diffusion of the Cu atoms along grain boundaries and toward the free surface depended on the misorientation of neighboring grains and the grain boundary energies \cite{Demas2017,jalilvand2019ectc}. The appearance of localized granular extrusion in Fig. \ref{fig:6}a was evidence that grain boundary diffusion creep and grain boundary sliding acted as the dominant mechanisms in this via. The protruded Cu and the grain boundaries did not seem to have a precise one-to-one correspondence, which further indicated the concurrence of grain boundary diffusion creep and grain boundary sliding. On the contrary, for the via in Fig. \ref{fig:6}b, except for a few small grains at the 6 o’clock and 11 o’clock positions at the peripheral, the majority of the via appeared in one contrast, suggesting that any grain boundaries, if existed in the interior of the via, had small misorientation (Fig. \ref{fig:6}e). Such a microstructure would have limited grain boundary diffusion in the inner region of the via to instead favor dislocation creep that involved dislocation glide and climb \cite{rupturebook2018}. This was consistent with the absence of mass accumulation in the interior of the via and the formation of a ring of extrusion around the via peripheral. The presence of some small grains at the via peripheral could accommodate grain boundary diffusion to drive extrusion, but the extent was expected to be small. In TSVs, interfacial diffusion and sliding at the via sidewall could drive via extrusion \cite{kumar2012interfacial}, but under the test condition in this study, no evidence of interfacial sliding was observed to cause the observed extrusion. In Figs. \ref{fig:6}c and \ref{fig:6}f, the SEM and FIB images were shown for another small-pitch via of the ring shape morphology that belonged to mode 2 of the distribution statistics. At an extrusion value of 1260nm, this via contained multiple grains with large contrast. Granular features (highlighted by arrows) were seen at the inner region of the via along with large protruded bumps at the peripheral. These microstructure and morphology features indicated the occurrence of both grain boundary diffusion creep, grain boundary sliding, and dislocation creep. Comparing the via in mode 2 (Fig. \ref{fig:6}c) with the via in mode 1 (Fig. \ref{fig:6}b) suggested that the concurrent operation of multiple deformation mechanisms under certain microstructure conditions could be a reason for the observed bimodal protrusion behavior. The smaller probability for a via to have such a stress and microstructure combination as that in Fig. \ref{fig:6}c resulted in the smaller number of vias in mode 2. Quantitative correlation of the extrusion morphology with microstructure would require detailed microstructure analysis by techniques such as electron backscatter diffraction, which would be the focus of future work. 

\begin{figure}[!b]
\centering
\includegraphics[width=0.48\textwidth]{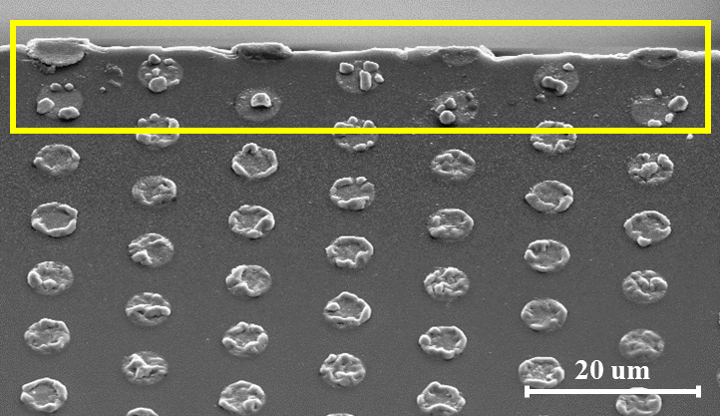}
\caption{SEM image at 45\degree tilt of TSV arrays at the free edge of the small-pitch sample with the edge vias highlighted.}
\label{fig:7} 
\end{figure}

Comparing the large-pitch and small-pitch samples, FEA showed that the overall stress level in the small-pitch vias was higher due to via proximity. The higher stress drove more plastic deformation, leading to the overall larger extrusion in the small-pitch vias. FEA also revealed that when the pitch was reduced, the stress enhancement was more pronounced around the via peripheral. Since the stress dependence of dislocation creep was stronger than that of diffusional creep and grain boundary sliding \cite{rupturebook2018,KASSNER2004fivepowerlawcreep}, higher stress in the small-pitch vias led to enhanced dislocation creep, especially near the via peripheral. As a result of their different stress levels, more annular morphology was observed in the small-pitch vias and more granular morphology was observed in the large-pitch vias. The effect of stress on extrusion morphology can also be seen by observing the vias at the free edge of the small-pitch sample. Highlighted in Fig. \ref{fig:7}, the row of vias closest to the free edge had less constraint and therefore was subjected to lower stress than the vias further away from the edge. In these vias, stress was insufficient to drive considerable dislocation creep and the deformation was mainly driven by grain boundary diffusion creep and grain boundary sliding. This was consistent with the appearance of sporadic bumps in the vias and the absence of continuous ring features. 

\begin{figure*}
\centering
\includegraphics[width=0.8\textwidth]{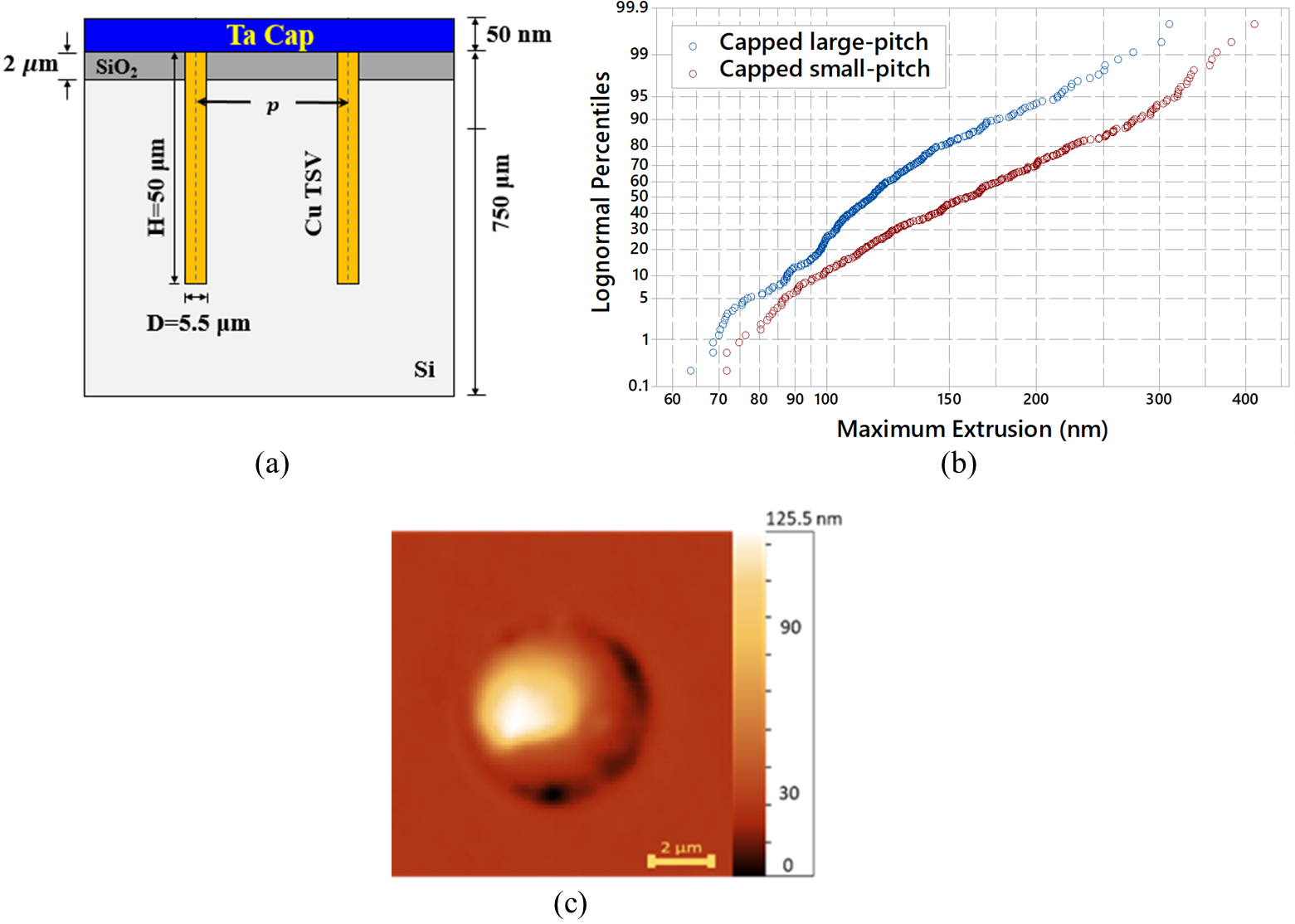}
\caption{(a) Cross-sectional view of a sample with the Ta cap layer. (b) Lognormal percentiles of the maximum extrusion for 300 Ta-capped large-pitch vias and 300 Ta-capped small-pitch vias. (c) WLI extrusion profile of  a capped small-pitch via with local dome shape extrusion morphology.}
\label{fig:8} 
\end{figure*}

Electroplated Cu TSVs were shown to have randomly orientated grains without a distinct texture \cite{DeMessemaeker2014,Spinella2016}. The small-pitch and the large-pitch samples were cut from the same wafer and were expected to have similar microstructural variance. The large statistical variation of extrusion seen in both samples (Fig. \ref{fig:2}) were attributed to the microstructural variance existed in the samples. In each extrusion mode,  the scale parameter of the granular morphology was larger than that of the annular morphology regardless of the pitch distance (Fig. \ref{fig:4}), suggesting the stronger influence of microstructural variation on the granular morphology.  Hence, the slightly larger extrusion spread of the large-pitch sample was related to the higher number of vias with granular extrusion, where deformation was predominantly accommodated by grain boundary diffusion. Yet, the difference in the  extrusion range was very small between the small- and large-pitch samples, suggesting that despite of the dominant mechanisms, the statistical variability of Cu microstructure still played the most critical role on the statistical spread of extrusion. This observation also suggested that if the microstructure were not optimized, using approaches based on physical design and post-plating annealing would not be able to control the statistics of extrusion, as seen by work reported in literature \cite{DeMessemaeker2013,DEWOL2011,Heryanto2012}.

\subsection{The effect of a cap layer in controlling extrusion and minimizing pitch effect}
\label{Discussion: cap layer}
Not only would the statistical variation of via extrusion be detrimental to the reliability of a 3D IC,  the change of extrusion due to pitch distance would also be undesirable in a 3D IC that contained vias with different pitch distances. Previously, we have demonstrated that a cap layer was effective in reducing the magnitude and variation of Cu extrusion \cite{JALILVAND2019scripta,jalilvand2019ectc}. The effect of a cap layer on the extrusion behaviors of vias with different pitch distances was investigated herein. A set of small-pitch and large-pitch samples identical to those described in section \ref{experiments} were coated with 50nm of tantalum (Ta) using a procedure reported elsewhere \cite{JALILVAND2019scripta} and then annealed. The schematic cross-sectional illustration of the TSVs coated with a Ta cap was illustrated in Fig. \ref{fig:8}a and these samples were called “capped small-pitch” and “capped large-pitch”, respectively. Following the same approach described in section \ref{experiments}, the maximum extrusion was determined for 300 vias each in the capped small- and large-pitch samples and the lognormal distribution plots were shown in Fig. \ref{fig:8}b.

In comparison to the case of the uncoated vias (Fig. \ref{fig:2}a), a considerable reduction of the maximum extrusion was observed in both capped samples. The $99^{th}$ percentile of the maximum extrusion was 311nm for the capped large-pitch vias and 412nm for the capped small-pitch vias, which were 69\% and 68\% reductions from their uncoated counterparts. The statistical spread of extrusion was considerably reduced for both the capped large- and small-pitch vias. An extrusion range of 247nm for large-pitch vias and 341nm for small-pitch vias were recorded, which were over 2.5-fold and 1.5-fold reductions from their uncoated counterparts. The difference between the small-pitch and large-pitch vias was also greatly reduced by the cap layer and the kinks in the distribution curves disappeared. Furthermore, morphological study on the capped vias from both samples revealed the complete disappearance of the annular and granular extrusion morphologies, which were replaced by a smooth “dome shape” morphology. An example of the dome shape morphology was shown in Fig. \ref{fig:8}c. 
The free top surface of an uncoated via served as vacancy sinks to facilitate the operation of time-dependent deformation mechanisms. The as-deposited Ta had been shown to form dense and continuous interface with Cu to effectively reduce the vacancy sources at the top of the via \cite{JALILVAND2019scripta}. The cap layer therefore would be effective in hindering the operation of both dislocation creep, grain boundary diffusion creep, and grain boundary sliding despite of the pitch distances to result in the observed reduction of extrusion and the disparities between the two samples. The Ta cap layer could also induce a compressive stress on the top of the via to oppose the motion of dislocations and reduce extrusion \cite{JALILVAND2019scripta}. The peaks of the domes were detected in varied locations within the vias, which could again be traced to the stochastic microstructure in Cu vias. However, as the statistical extrusion spread in the uncoated vias was driven by microstructure accommodated operation of the deformation mechanisms, suppressing the deformation in the capped vias led to a reduced spread of extrusion in both samples despite of the microstructure variations. In this case, the larger variation observed in the capped small-pitch vias was mainly attributed to the higher stress caused by via proximity. The ability of the cap layer to effectively lower the magnitude of extrusion and more importantly to reduce the spread of extrusion made it a promising approach to mitigate the extrusion problem in 3D ICs employing high density TSVs with different pitch distances.

\section{Conclusion}
\label{Conclusion}
In this work, we investigated the effect of pitch distance on the magnitude, statistical variation, and morphology of extrusion in Cu TSVs. Smaller pitch distance induced larger stresses in the via, particularly at the via peripherals, and consequently resulted in larger magnitude of extrusion. The variation of extrusion was fundamentally related to the stochastic nature of Cu microstructure and was not significantly affected by the pitch distance. Two classes of extrusion morphology, the granular morphology and the annular morphology, were observed and attributed to the operation and relative dominance of different deformation mechanisms under the combined effect of stress and microstructure. The granular morphology was more populated in the large-pitch vias due to the dominance of creep and grain boundary sliding accommodated by grain boundary diffusion. The annular morphology was more populated in the small-pitch vias, which was attributed to the stress enhanced dislocation creep. The concurrent contribution of multiple dominant mechanisms in some vias with certain microstructure and stress states also led to the bimodal extrusion distribution observed in both samples. Finally, it was shown that the magnitude and more importantly the statistical spread of extrusion in Cu TSVs with different pitch distance could be effectively reduced by a Ta cap layer.

\section*{Acknowledgment}

The authors would like to thank SEMATECH for the TSV samples. The  financial support by UCF Startup fund is also acknowledged.

\ifCLASSOPTIONcaptionsoff
  \newpage
\fi

\bibliographystyle{IEEEtran}
\bibliography{References.bib}

\end{document}